# Information Hiding Techniques: A Tutorial Review

**Sabu M Thampi**
Assistant Professor
Department of Computer Science & Engineering
LBS College of Engineering, Kasaragod
Kerala- 671542, S.India
smtlbs@yahoo.co.in

## Abstract
*The purpose of this tutorial is to present an overview of various information hiding techniques. A brief history of steganography is provided along with techniques that were used to hide information. Text, image and audio based information hiding techniques are discussed. This paper also provides a basic introduction to digital watermarking.*

## 1. History of Information Hiding

The idea of communicating secretly is as old as communication itself. In this section, we briefly discuss the historical development of information hiding techniques such as steganography/ watermarking.

Early steganography was messy. Before phones, before mail, before horses, messages were sent on foot. If you wanted to hide a message, you had two choices: have the messenger memorize it, or hide it on the messenger.

While information hiding techniques have received a tremendous attention recently, its application goes back to Greek times. According to Greek historian Herodotus, the famous Greek tyrant Histiaeus, while in prison, used unusual method to send message to his son-in-law. He shaved the head of a slave to tattoo a message on his scalp. Histiaeus then waited until the hair grew back on slave's head prior to sending him off to his son-in-law.

The second story also came from Herodotus, which claims that a soldier named Demeratus needed to send a message to Sparta that Xerxes intended to invade Greece. Back then, the writing medium was written on wax-covered tablet. Demeratus removed the wax from the tablet, wrote the secret message on the underlying wood, recovered the tablet with wax to make it appear as a blank tablet and finally sent the document without being detected.

Invisible inks have always been a popular method of steganography. Ancient Romans used to write between lines using invisible inks based on readily available substances such as fruit juices, urine and milk. When heated, the invisible inks would darken, and become legible. Ovid in his "Art of Love" suggests using milk to write invisibly. Later chemically affected sympathetic inks were developed. Invisible inks were used as recently as World War II. Modern invisible inks fluoresce under ultraviolet light and are used as anti-counterfeit devices. For example, "VOID" is printed on checks and other official documents in an ink that appears under the strong ultraviolet light used for photocopies.

The monk Johannes Trithemius, considered one of the founders of modern cryptography, had ingenuity in spades. His three volume work *Steganographia,* written around 1500, describes an extensive system for concealing secret messages within innocuous texts. On its surface, the book seems to be a magical text, and the initial reaction in the 16th century was so strong that *Steganographia* was only circulated privately until publication in 1606. But less than five years ago, Jim Reeds of AT&T Labs deciphered mysterious codes in the third volume, showing that Trithemius' work is more a treatise on cryptology than demonology. Reeds' fascinating account of the code breaking process is quite readable.

One of Trithemius' schemes was to conceal messages in long invocations of the names of angels, with the secret message appearing as a pattern of letters within the words. For example, as every other letter in every other word:

**p**adiel a**p**o**r**s**y** mesarpon **o**m**eu**as peludyn m**a**l**pre**a**x**o

which reveals "prymus apex."

Another clever invention in *Steganographia* was the "Ave Maria" cipher. The book contains a series of tables, each of which has a list of words, one per letter. To code a message, the message letters are replaced by the corresponding words. If the tables are used in order, one table per letter, then the coded message will appear to be an innocent prayer.

The earliest actual book on steganography was a four hundred page work written by Gaspari Schott in 1665 and called *Steganographica*. Although most of the ideas came from Trithemius, it was a start.

Further development in the field occurred in 1883, with the publication of Auguste Kerchoffs' *Cryptographie militaire*. Although this work was mostly about cryptography, it describes some principles that are worth keeping in mind when designing a new steganographic system.

But it was during the twentieth century that steganography truly flowered. An example of this comes from early in the century, during the Boer War. The British employed Lord Robert Baden-Powell founder of Boy Scout movement as a scout. His job was to mark the positions of Boer artillery bases. To ensure he was not suspected by the Boers, if he was caught, he would mark his maps into drawings of butterflies. Appearing innocent to a casual observer, certain markings on the wings were actually the positions of the enemy military installations.

During World War II, null ciphers (unencrypted message) were used to hide secret messages. The null cipher, which often appeared to be innocent message about ordinary occurrences, would not alert suspicion, and would thus not be intercepted. For example, the following message was sent by German spy during WWII.

*Apparently neutral's protest is thoroughly discounted and ignored. Isman hard hit. Blockade issue affects pretext for embargo on by-products, ejecting suets and vegetable oils.*

Decoding this message by taking the second letter in each word reveals the following secret message.

*Pershing sails from NY June 1.*

With the advent of photography, microfilm was created as a way to store a large amount of information in a very small space. In both world wars, the Germans used "microdots" to hide information, a technique which J. Edgar Hoover called "the enemy's masterpiece of espionage." A secret message was photographed, reduced to the size of a printed period, and then pasted into an innocuous cover message, magazine, or newspaper. The Americans caught on only when tipped by a double agent: "Watch out for the dots -- lots and lots of little dots."

A whole other branch of steganography, Linguistic steganography, consists of linguistic of language form of hidden writing. These are the "semagrams" and the "open code". A semagram is a secret message that is not in a written form. For example, a system can use long blades of grass in a picture as dashes in Morse code, with short blades for dots. Open codes are illusions or code words. In World War 1, for example, German spies used fake orders for cigars to represent various types of British warships – cruisers and destroyers. Thus 500 cigars needed in Portsmouth meant that five cruisers were in Portsmouth.

**Fingerprints** are characteristics of an object that tend to distinguish it from other similar objects. They enable the owner to trace authorized users distributing them illegally. Digital fingerprinting produces a metafile that describe the contents of the source file. In the case of encrypted satellite television broadcasting, for instance, users could be issued a set of keys to decrypt the video streams and the television station could insert fingerprint bits into each packet of the traffic to detect unauthorized uses. If a group of users give their subset of keys to unauthorized people (so that they can also decrypt the traffic) at least one of the key donors can be traced when the unauthorized decoder is captured.

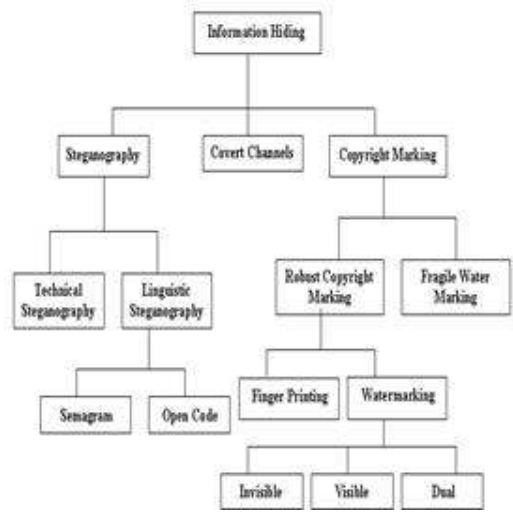

**Figure 1 - Information Hiding Techniques**

A **covert channel** could be defined as a communications channel that transfers some kind of information using a method originally not intended to transfer this kind of information. Observers are unaware that a covert message is being communicated. Only the sender and recipient of the message notice it.

With the computer age, information hiding has been given a marvelous boost. We are sure to see a great expansion of steganographical techniques in the coming years.

## 2. What is Steganography and why is it important?

Steganography or Stego as it often referred to in the IT community, literally means, "Covered writing" which is derived from the Greek language. Steganography is defined as follows,

"Steganography is the art and science of communicating in a way which hides the existence of the communication. The goal of Steganography is to hide messages inside other harmless messages in a way that does not allow any enemy to even detect that there is a second message present".

In a digital world, Steganography and cryptography are both intended to protect information from unwanted parties. Both Steganography and Cryptography are excellent means by which to accomplish this but neither technology alone is perfect and both can be broken. It is for this reason that most experts would suggest using both to add multiple layers of security.

**Steganography Vs Cryptography**

The term Steganography means, "cover writing" whereas cryptography means "secret writing". Cryptography is the study of methods of sending messages in distinct form so that only the intended recipients can remove the disguise and read the message. The message we want to send is called plain text and disguised message is called cipher text. The process of converting a plain text to a cipher text is called enciphering or encryption, and the reverse process is called deciphering or decryption. Encryption protects contents during the transmission of the data from sender to receiver. However, after receipt and subsequent decryption, the data is no longer protected and is the clear. Steganography hides messages in plain sight rather than encrypting the message; it is embedded in the data (that has to be protected) and doesn't require secret transmission. The message is carried inside data.

Steganography can be used in a large amount of data formats in the digital world of today. The most popular data formats are .bmp, .doc, .gif, .jpeg, .mp3, .txt and .wav. Steganographic technologies are a very important part of the future of Internet security and privacy on open systems such as Internet.

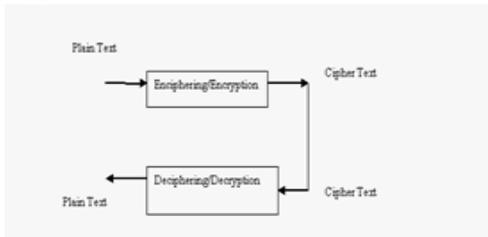

**Figure 2a : Cryptography**

Steganographic research is primarily driven by the lack of strength in the cryptographic systems on their own and the desire to have complete secrecy in an open-systems environment. Many Governments have created laws that either limit the strength of cryptosystems or prohibit them completely. This unfortunately leaves the majority of the Internet community either with relatively weak and a lot of the times breakable encryption algorithms or none at all. This is where Steganography comes in. Steganography can be used to hide important data inside another file so that only the parties intended to get the message even knows a secret message exists. It is a good practice to use Cryptography and Steganography together.

Neither Steganography nor Cryptography is considered "turnkey solutions" to open systems privacy, but using both technologies together can provide a very acceptable amount of privacy for anyone connecting to and communicating over these systems.

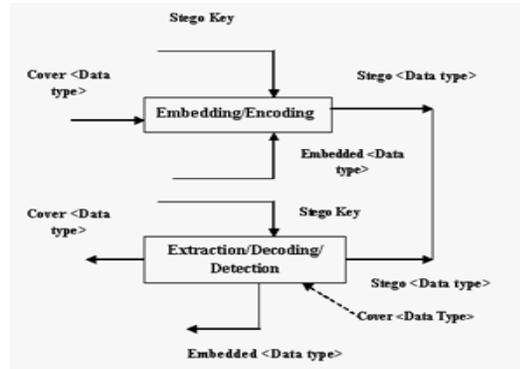

**Figure 2b: Steganography**

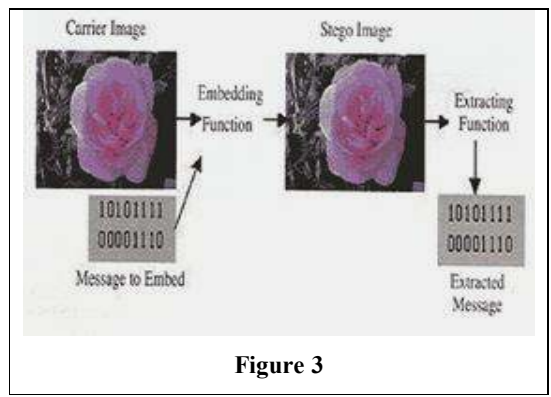

**Figure 3**

## 3. A Detailed Look at Steganography

In this section we will discuss Steganography at length. We will start by looking at the different types of Steganography generally used in practice today along with some of the other principles that are used in Steganography.

To start, let's look at what a theoretically perfect secret communication (Steganography) would consist of. To illustrate this concept, we will use three fictitious characters named Amy, Bret and Crystal. Amy wants to send a secret message (M) to Bret using a random (R) harmless message to create a cover (C), which can be sent to Bret without raising suspicion. Amy then changes the cover message (C) to a stego-object (S) by embedding the secret message (M) into the cover message (C) by using a stego-key (K). Amy should then be able to send the stego-object (S) to Bret without being detected by Crystal. Bret will then be able to read the secret message (M) because he knows the stego-key (K) used to embed it into the cover message (C).

In order to embed secret data into a cover message, the cover must contain a sufficient amount of redundant data or noise. This is because in the embedding process Steganography actually replaces redundant data with the secret message. This limits the types of data that we can use with Steganography.

There are basically three types of steganographic protocols used. They are Pure Steganography, Secret Key Steganography, and Public Key Steganography.

Pure Steganography is defined as a steganographic system that does not require the exchange of a cipher such as a stego-key. This method of Steganography is the least secure means by which to communicate secretly because the sender and receiver can rely only upon the presumption that no other parties are aware of this secret message.

Secret Key Steganography is defined as a Steganographic system that requires the exchange of a secret key (stego-key) prior to communication. Secret Key Steganography takes a cover message and embeds the secret message inside of it using a secret key (stego-key). Only the parties who know the secret key can reverse the process and read the secret message. Unlike Pure Steganography where a perceived invisible communication channel is present, Secret Key Steganography exchanges a stego-key, which makes it more susceptible to interception. The benefit to Secret Key Steganography is even if it is intercepted; only parties who know the secret key can extract the secret message.

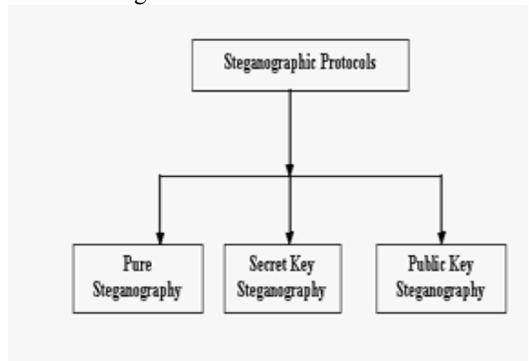

**Figure 4: Steganographic Protocols**

Public Key Steganography takes the concepts from Public Key Cryptography as explained below. Public Key Steganography is defined as a steganographic system that uses a public key and a private key to secure the communication between the parties wanting to communicate secretly. The sender will use the public key during the encoding process and only the private key, which has a direst mathematical relationship with the public key, can decipher the secret message. Public Key Steganography provides a more robust way of implementing a steganographic system because it can utilise a much more robust and researched technology in Public Key Cryptography.

Throughout the history different media types have been used to hide information. With advancements in computer industry this number is only increasing. Some of the media types are computer file system, transmission protocols, audio files, text files and images. A brief introduction for encoding messages in various media types is given below.

> **Kerchoff's Principle**
>
> *The security of the system has to be based on the assumption that the enemy has full knowledge of the design and implementation details of the steganographic system. The only missing information for the enemy is a short, easily exchangeable random number sequence, the secret key. Without this secret key, the enemy should not have the chance to even suspect that on an observed communication channel, hidden communication is taking place.*

### 3.2 Computer File System
Where it stores normal data, a computer file system can also be used to hide information between innocent files. For example a hard drive while showing the visible partition to a computer user may contain hidden partitions that can carry hidden files inside them. For example sfspatch is a kernel patch, which introduces module support for the steganographic file on a Linux machine. Sfspatch employs encryption along with steganographic techniques to hide information on the disk so it is not visible to a casual user.

FAT 16 system on Microsoft Windows hosts allocate 32 kilobytes of disk space to each file. If the file size is only a few kilobytes, the rest of the space can be used to hide information.

### 3.3 Transmission Protocol
Covert channels can be established using the control data, timing properties of transmission or of the user data. In this approach it is very difficult or almost impossible to prove the existence of covert channels, because the information is stripped off at the receiver. But if the information is hidden using user data, it remains on the hard disk until it is specifically deleted. Thus network systems can be utilized in cryptography to establish hidden channels of communications.

Transmission Control Protocol (TCP) and Internet Protocol (IP) are some of the few protocols that can be used to hide information inside certain header fields.

Some TCP/IP fields are either changed or stripped off by packet filtering mechanisms or through fragment re-assembly. However, there are fields that are less likely to change or altered. These fields include: Identification field, Sequence Number field and Acknowledge Sequence Number field.

- **Hiding information within an IP header**

The Identification field within an IP header provides network devices with a unique number to identify packets that may require reassembly. As presented by Neil F. Johnson in INFS 762 class at George Mason University (GMU), replacing the identification field with the numerical ASCII representation of the character to be encoded provides an easy way to hide information within this field. In his example, Johnson selects an unsigned integer to be transmitted as the identification field. The ASCII value of this integer can be achieved by dividing the integer by 256.

At the transmitting end client host construct a packet to include the desired identification number along with source and destination address. In this example I have chosen 18432, 18688, 17408 and 17664 as the four identification field values for the four IP packets. This process is depicted in Table 1. Once the ASCII value of the identification field is calculated at the destination, the decoded message is found to be the word "HIDE."

| 00 01 02 03 04 05 06 07 08 09 10 11 12 13 14 15 16 17 18 19 20 21 22 23 24 25 26 27 28 29 30 31 |||||
|---|---|---|---|---|
| Version | IHL | TOS | Total length ||
| Identification ||| Flags | Fragment offset |
| TTL || Protocol | Header checksum ||
| Source IP address |||||
| Destination IP address |||||
| Options |||| Padding |

**Table 1: The IP Header**

Similar techniques can also be used to encode information in the Sequence Number field of a TCP packet.

- **Session Layer**

An Open Systems Interconnection Reference model (OSI) uses packet structures to send information across the network from one layer to another as well as from one network terminal to another. A network packet consists of packet headers, user data and packet trailers. All the packets sent across the network have the same packet structure.

The session layer allows two machines to establish sessions over the network. These sessions allow ordinary data transfer plus enhanced services for some applications. This function is achieved via software that can "mount" remote discs on a local machine. Richard Popa, who has conducted research in this area at the University of Timisoara, has described the following scheme that can be used to establish covert communication channel:

*"Suppose we have two files on the disk of Alice, Bob can read one of them. If he reads the first file then Alice records a zero and if he reads the second file she records a one."*

| Sending Host | Receiving Host |
|---|---|
| 21:55:35.763195 IP www.source.com.80 > www.destination.com.1444: S 788196017:788196017(0) ack 2449684694 win 32640 (ttl 64, id 18432) | .....(ttl 64, id 18432/256 ) ASCII 72 = H |
| 21:55:34.754811 IP www.source.com.80 > www.destination.com.1442: S 1093146807:1093146807(0) ack 2449313332 win 64240 (ttl 64, id 18688) | .....(ttl 64, id 18688/256 ) ASCII 73 = I |
| 21:55:33.711164 IP www.source.com.80 > www.destination.com.1439: S 1092753896:1092753896(0) ack 2448919278 win 64240 (ttl 64, id 17408) | .....(ttl 64, id 17408/256 ) ASCII 68 = D |
| 21:55:32.304123 IP www.source.com 80 > www.destination.co.1436: S 1092243953:1092243953(0) ack 2448416550 win 64240 (ttl 64, id 17664) | .....(ttl 64, id 17664/256 ) ASCII 69 = E |

**Table 2: Hiding data in the identification field**

The fact that Oscar can see this traffic should not arouse his suspicion, since it is irrelevant to him that Bob reads one file rather than another.

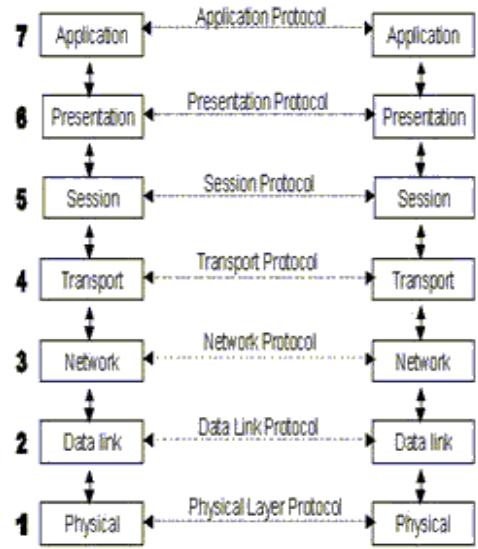

**Figure 5: Open Systems Interconnection Reference model (OSI)**

### 3.4 Encoding Secret Messages in Text

Encoding secret messages in text can be a very challenging task. This is because text files have a very small amount of redundant data to replace with a secret message. Another drawback is the ease of which text based Steganography can be altered by an unwanted parties by just changing the text itself or reformatting the text to some other form (from .txt to .doc, etc.). There are numerous methods by which to accomplish text based Steganography. These methods are: Open space method, Syntactic methods and Semantic methods (Figure 4).

### 3.4.1 Open Space Methods

There are couple ways to employ the open space in text files to encode the information. This method works because to a casual reader one extra space at the end of line or an extra space between two words does not prompt abnormality. However, open space methods are only useful with ASCII format.

- Inter-sentence space method encodes a "0" by adding a single space after a period in English prose. Adding two spaces would encode a "1". This method works, but requires a large amount of data to hide only little information. Also many word processing tools automatically correct the spaces between sentences.

Line-shift encoding involves actually shifting each line of text vertically up or down by as little 3 centimeters. Depending on whether the line was up or down from the stationary line would equate to a value that would or could be encoded into a secret message.

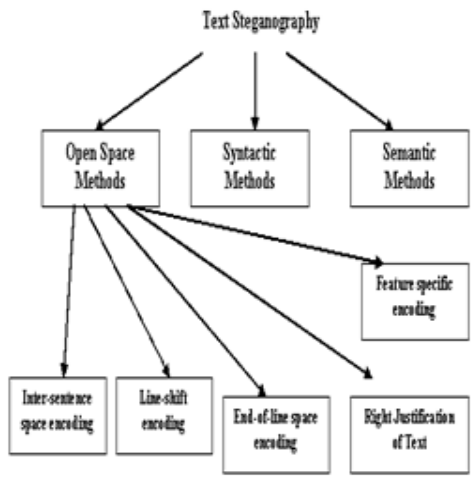

**Figure 6: Steganography in Text**

- End-of-line space method exploits white space at the end of each line. Data encoded using a predetermined number of spaces at the end of each line. For example two spaces will encode one bit, four spaces will encode two bits and eight spaces will encode three bits and so on. This technique works better than the inter-space method, because increasing the number of spaces can hide more data.

- Right-justification of text can also be used to encode data within text files. Calculating and controlling the spaces between words encode data in innocent text files. One space between words represents a "0" and two spaces represent a "1". However, this approach makes it difficult to decode the data as it becomes impossible to distinguish a single innocent space form an encoded one. For this purpose another technique based on Manchester coding is used. Hence "01" is interpreted as "1" and "10" is interpreted as "0". Whereas "00" and "11" are considered the null bit strings.

- Feature specific encoding involves encoding secret messages into formatted text by changing certain text attributes such as vertical/horizontal length of letters such as b, d, T, etc. This is by far the hardest text encoding method to intercept as each type of formatted text has a large amount of features that can be used for encoding the secret message.

All of the above text based encoding methods require either the original file or the knowledge of the original files formatting to be able to decode the secret message.

### 3.4.2 Syntactic Methods

Syntactic methods exploit the use of punctuation and structure of text to hide data without scientifically altering the meaning of the message. For example the two phrases *"bread, butter, and milk"* and *"bread, butter and milk"* are grammatically correct but differ in the use of comma. One can employ this structure alternatively in a text message to represent either a "1" of one method is used and to represent a "0" if the other method is employed.

### 3.4.3 Semantic Methods

Semantic methods assign two synonyms a primary or secondary value. These values are then translated into binary "1" or "0". For example the word "big" is assigned a primary and the word "large" is assigned secondary. Therefore, decoding a message would translate the use of primary to be "1" and secondary to a "0". The problem in this approach is that replacement of synonyms may change the meaning or structure of the sentence. For example calling someone "cool" has a different meaning than calling him or her "chilly".

Our tool analysis for text file Steganography will cover a product namely "SNOW", that makes use of *end of lines of text* for encoding messages.

### 3.4.4 Concealing Messages in Text Files: SNOW

Snow is a free for non-commercial use program available at http://www.darkside.com.au/snow and is authored by Matthew Kwan.

The encoding scheme used by **snow** relies on the fact that spaces and tabs (known as *whitespace*), when appearing at the end of lines, are invisible when displayed in pretty well all text viewing programs. This allows messages to be hidden in ASCII text without affecting the text's visual representation. And since trailing spaces and tabs occasionally occur naturally, their existence should not be sufficient to immediately alert an observer who stumbles across them.

The **snow** program runs in two modes - message concealment, and message extraction.

During concealment, the following steps are taken.

Message -> optional compression -> optional encryption -> concealment in text

Extraction reverses the process.
Extract data from text -> optional decryption -> optional uncompression -> message
The program has the handy ability to tell you how much data it can fit in the desired cover file: Issuing the command "snow –S cover.txt" produces the following output: File has storage capacity of between 1763 and 2012 bits. Our embedded data, my secret.txt is only 107 bytes, so we know it will easily fit within this cover file. If space is of concern, the program also offers a flag to compress the data, however the author notes that "if the data is not text, or of there is lot of data, the use of the built-in compression is not recommended", and suggests that the user pre-compress the data with more robust compression tools such as winzip.

Snow also provides the ability to encrypt the data to be hidden with a password-protected key. It uses the author's own ICE encryption protocol, allowing for passwords or pass phrases of up to 1170 characters. ICE stands for *Information Concealment Engine*. It is a 64-bit private key block cipher, in the tradition of DES. However, unlike DES, it was designed to be secure against differential and linear cryptanalysis, and has no key complementation weaknesses or weak keys. In addition, its key size can be any multiple of 64 bits, whereas the DES key is limited to 56 bits. The ICE algorithm is public domain, and source code can be downloaded.

To embed our secret data in our file we issue the command:

snow –C –f mysecret.txt –p mypassword cover.txt stego.txt

This command compresses the message contained in mysecret.txt, encrypts it with the password-protected key using "mypassword" and embeds it in cover.txt, creating the stego file called stego.txt. The output from the command informs the user that it compressed the original message by 41.87% and that the message used approximately 25.14% of the available space in the cover file.

The extraction process is just as straightforward, issuing the command:

snow –C –p mypassword stego.txt

will output the contents of our secret message file to open.

The stego.txt, when opened with a common text editor, like Microsoft Word, looks identical to the original, despite having gained 655 bytes in size with the addition of the secret message. Also, by telling Microsoft Word to show special formatting marks, one can easily see the inserted tabs and spaces in the stego document. Due to the presence of strong encryption scheme and without knowing the password the attacker can not extract the hidden message.

### 3.5 Data Hiding in the Graphic Files
Coding secret messages in digital images is by far the most widely used of all methods in the digital world of today. This is because it can take advantage of the limited power of the human visual system (HVS). Almost any plain text, cipher text, image and any other media that can be encoded into a bit stream can be hidden in a digital image. With the continued growth of strong graphics power in computers and the research being put into image based Steganography, this field will continue to grow at a very rapid pace.

To a computer, an image is an array of numbers that represent light intensities at various points, or pixels. These pixels make up the images raster data. When dealing with digital images for use with Steganography, 8-bit and 24-bit per pixel image files are typical. Both have advantages and disadvantages, as we will explain below. 8-bit images are a great format to use because of their relatively small size. The drawback is that only 256 possible colors can be used which can be a potential problem during encoding. Usually a gray scale color palette is used when dealing with 8-bit images such as (.GIF) because its gradual change in color will be harder to detect after the image has been encoded with the secret message. 24-bit images offer much more flexibility when used for Steganography. The large numbers of colors (over 16 million) that can be used go well beyond the HVS, which makes it very hard to detect once a secret message, has bee encoded. The other benefit is that a much larger amount of hidden data can be encoded into 24-bit digital image as opposed to an 8-bit digital image. The one major drawback to 24-bit digital images is their large size (usually in MB) makes them more suspect than the much smaller 8-bit digital images (usually in KB) when sent over an open system such as Internet. Digital image compression (lossy compression – jpeg) is a good solution to large digital images such as 24-bit images mentioned earlier.

Information can be hidden many different ways in images. Straight message insertion can be done, which will simply encode every bit of information in the image. More complex encoding can be done to embed the message only in ``noisy'' areas of the image that will attract less attention. The message may also be scattered randomly throughout the cover image. The most common approaches to information hiding in images are:

- Least significant bit (LSB) insertion
- Masking and filtering techniques
- Transformations

Each of these can be applied to various images, with varying degrees of success. Each of them suffers to varying degrees from operations performed on images, such as cropping, or resolution decrementing, or decreases in the colour depth.

- **Least Significant Bit Insertion**

The least significant bit insertion method is probably the most well-known image steganography technique. It is a common, simple approach to embedding information in a graphical image file. Unfortunately, it is extremely vulnerable to attacks, such as image manipulation. A simple conversion from a GIF or BMP format to a lossy compression format such as JPEG can destroy the hidden information in the image.

When applying LSB techniques to each byte of a 24-bit image, three bits can be encoded into each pixel. (As each pixel is represented by three bytes.) Any changes in the pixel bits will be indiscernible to the human eye. For example, the letter A can be hidden in three pixels. Assume the original three pixels are represented by the three 24-bit words below:

( 00100111 11101001 11001000 ) ( 00100111 11001000 11101001 ) ( 11001000 00100111 11101001 )

The binary value for the letter A is (10000011). Inserting the binary value of A into the three pixels, starting from the top left byte, would result in:

(00100111 1110100*0* 11001000)
(00100110 11001000 1110100*0*)
(11001000 00100111 11101001)

The emphasised bits are the only bits that actually changed. The main advantage of LSB insertion is that data can be hidden in the least and second to least bits and still the human eye would be unable to notice it.

When using LSB techniques on 8-bit images, more care needs to be taken, as 8-bit formats are not as forgiving to data changes as 24-bit formats are. Care needs to be taken in the selection of the cover image, so that changes to the data will not be visible in the stego-image. Commonly known images, (such as famous paintings, like the *Mona Lisa*) should be avoided. In fact, a simple picture of your dog would be quite sufficient.

When modifying the LSB bits in 8-bit images, the pointers to entries in the palette are changed. It is important to remember that a change of even one bit could mean the difference between a shade of red and a shade of blue. Such a change would be immediately noticeable on the displayed image, and is thus unacceptable. For this reason, data-hiding experts recommend using grey-scale palettes, where the differences between shades is not as pronounced.

- **Masking and Filtering**

Masking and filtering techniques hide information by marking an image in a manner similar to paper watermarks. Because watermarking techniques are more integrated into the image, they may be applied without fear of image destruction from lossy compression. By covering, or masking a faint but perceptible signal with another to make the first non-perceptible, we exploit the fact that the human visual system cannot detect slight changes in certain temporal domains of the image.

Masking techniques are more suitable for use in lossy JPEG images than LSB insertion because of their relative immunity to image operations such as compression and cropping.

- **Transformations**

Transform Domain tools utilize an algorithm such as the Discrete Cosine Transformation (DCT) or wavelet transformation to hide information in significant areas of the image. The JPEG image format uses a discrete cosine transformation (DCT) to transform successive 8x8 pixel blocks of the image into 64 DCT coefficients each. The least-significant bits of the quantized DCT coefficients are used as redundant bits into which the hidden message is embedded. The modification of a single DCT coefficient affects all 64-image pixels. For that reason, there are no known visual attacks against the JPEG image format.

Stego-tools which utilize one of the many transform domain techniques are more robust, have a higher resilience to attacks against the stego-image such as compression, cropping and image processing. As of this writing all of the stego-tools which can manipulate JPEG images are transform domain tools such as;

Jpeg-Jsteg, JPHide, Outguess, PictureMarc and SysCop.

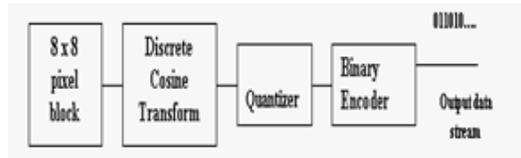

Figure 7: Block Diagram of JPEG image compression

### 3.5.1 Pixel Calculator

There are a number of easy to use tools available on the Internet to hide information in image files. To better understand and appreciate some of the processes used by these tools some understanding of digital image processing becomes essential. Written by Steve Tenimoto and his team at University of Washington, Pixel Calculator is a very interesting tool to understand digital images. Pixel Calculator also provides a neat feature of achieving some very basic Steganography.

Pixel calculator is equipped with two basic tools. A zooming tool is provided to learn the exact pixel value by zooming into an image until the pixel values are visible. A calculator tool is then used to change or modify pixel values. Learning the pixel values and changing them using the calculator is the key in hiding the information inside an image. Figure:6 depicts the image file I used as cover to hide the information.

I used the zooming tool to find an area of interest where neighboring pixel values are close to each other. These values are visible in Figure 8. Using the calculator tool, I started to replace the pixel values in the image with a magic number, say 90. I repeated this process until I was done typing the hidden message.

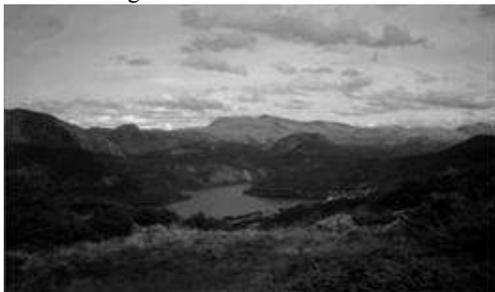

Figure 9 shows the image with a hidden message on the left hand side and the decoded message at the right hand side. The red circle on top of the mountain peak represents the area where the message is hidden. To decode the message, calculator tool is used again to convert bits lower than the "magic number" to "0" whereas, the higher ones are converted to "255." This process converts rest of the image black and white, while revealing the hidden message in gray color.

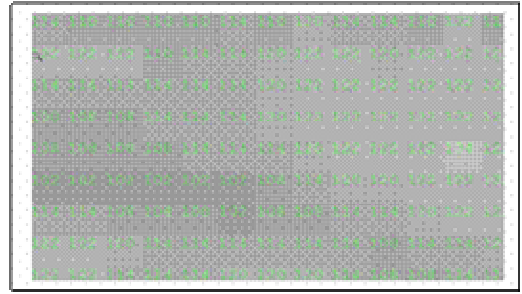

Figure 8: Cover Image and Uniform Pixel Area

### 3.5.2 Concealing Messages in JPEG Image Files: Jsteg

Jsteg hides the data inside images stored in the JFIF format of the JPEG standard. It was believed that this type of steganography was impossible, or at least infeasible, since the JPEG standard uses lossy encoding to compress its data. Any hidden data would be overwhelmed by the noise. The trick used by this steganographic implementation is to recognize that JPEG encoding is split into lossy and non-lossy stages. The lossy stages use a discrete cosine transform and a quantization step to compress the image data; the non-lossy stage then uses Huffman coding to further compress the image data. As such, we can insert the steganographic data into the image data between those two steps and not risk corruption.

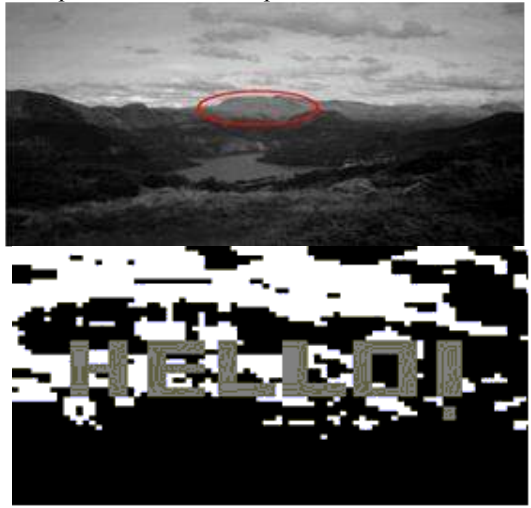

Figure 9: Encoded Picture and Decoded Message

To compile the package, simply follows the steps given:

To inject a data file into a JPEG/JFIF image, simply add the option "-steg filename" to the "cjpeg" command line. If the data file is too large for the image, "cjpeg" will inform you. At this point, you can compress the data file, increase the quality of the image (thereby increasing image size), or try a different image.

Extraction of a data file works similarly. The "-steg filename" option to "djpeg" writes the steganographic data to the file, wiping out its previous contents. Usually, the decoded image sent to standard output is redirected to "/dev/null".'

## 3.6 Data Hiding in Audio Files

Encoding secret messages in audio is the most challenging technique to use when dealing with Steganography. This is because the human auditory system (HAS) has such a dynamic range that it can listen over. The HAS perceives over a range of power greater than one billion to one and a range of frequencies greater than one thousand to one. Sensitivity to additive random noise is also acute. Perturbations in a sound file can be detected as low as one part in ten million. However there are some "holes" available in this perspective range where data may be hidden. While the HAS has a large dynamic range, it often has a fairly small differential range. As a result, loud sounds tend to mask out quiet sounds. There are some environmental distortions so common as to be ignored by the listener in most cases.

There are two concepts to consider before choosing an encoding technique for audio. They are the digital format of the audio and the transmission medium of the audio.

There are three main digital audio formats typically in use. They are Sample Quantization, Temporal Sampling Rate and Perceptual Sampling.

- Sample Quantization which is a 16-bit linear sampling architecture used by popular audio formats such as .WAV and .AIFF.
- Temporal Sampling uses selectable frequencies (8 kHz, 9.6 kHz, 10 kHz, 12 kHz, 16 kHz, 22.05 kHz and 44.1 kHz.) to sample the audio. Sampling rate puts an upper bound on the usable portion of the frequency range. Generally, the higher the sampling rate is, the higher the usable data space gets.
- Perceptual Sampling format changes the statistics of the audio drastically by encoding only the parts the listener perceives, thus maintaining the sound but changing the signal. This format is used by the most popular digital audio on the Internet today in ISO MPEG (MP3).

Transmission medium (path the audio takes from sender to receiver) must also be considered when encoding secret messages in audio. The four transmission mediums are discussed below.

- Digital end-to-end environment: If a sound file is copied directly from machine to machine, but never modified, then it will go through this environment. As a result, the sampling will be exactly the same between the encoder and decoder. Very little constraints are put on data hiding in this environment.
- Increased/decreased resampling environment: In this environment, a signal is resampled to a higher or lower sampling rate, but remains digital throughout. Although the absolute magnitude and phase of most of the signal are preserved, the temporal characteristics of the signal are changed.
- Analog transmission and resampling: This occurs when a signal is converted to an analog state, played on a relatively clean analog line, and resampled. Absolute signal magnitude, sample quantisation and temporal sampling rate are not preserved. In general, phase will be preserved.
- "Over the air" environment: This occurs when the signal is ``played into the air'' and ``resampled with a microphone''. The signal will be subjected to possible unknown nonlinear modifications causing phase changes, amplitude changes, drifting of different frequency components, echoes, etc.

The signal representation and transmission environment both need to be considered when choosing a data-hiding method.

### 3.6.1 Methods of Audio Data Hiding

We now need to consider some methods of audio data hiding.

➢ In **low-bit encoding** data is embedded by replacing the Least Significant Bit (LSB) of each sampling point by a coded binary string. This results in a large amount of data that can be encoded in a single audio file. For example if the ideal noiseless channel capacity is I Kbps then the bit rate will be 8 Kbps given an 8 kHz sampled sequence. While the simplest way to hide data in the audio files, low-bit encoding scheme can be destroyed by the channel noise and re-sampling.
➢ **Phase coding** when it can be used has proven to be most effective coding techniques in terms of signal to noise ratio. In this method the phase of the original audio signal is replaced with the reference phase of the data to be hidden. It is discovered that a channel capacity of approximately 8 bps can be achieved by allocating 128 frequency slots per bit with a little background noise. The procedure for phase coding is as follows:

- The original sound sequence is broken into a series of N short segments.
- A discrete Fourier transform (DFT) is applied to each segment, to break create a matrix of the phase and magnitude.
- The phase difference between each adjacent segment is calculated.
- For segment S0, the first segment, an artificial absolute phase p0 is created.

- For all other segments, new phase frames are created.
- The new phase and original magnitude are combined to get a new segment, Sn.
- Finally, the new segments are concatenated to create the encoded output.

For the decoding process, the synchronisation of the sequence is done before the decoding. The length of the segment, the DFT points, and the data interval must be known at the receiver. The value of the underlying phase of the first segment is detected as 0 or 1, which represents the coded binary string

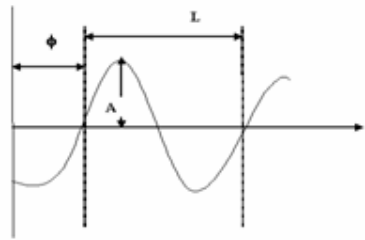

**Figure 10: A sinusoidal Function*, characterized by a period (L), an amplitude (A) and a phase(ϕ)**

*Note: The length of the cycle, L is known as the period of the function. The amplitude is the size of the variation – the height of a peak or depth of a trough. The phase is the position of the start of cycle, relative to some reference point (e.g., the origin) A sine function has $\phi =0$, whereas a cosine function has $\phi=\pi/2$*

- Modern steganographic systems use **spread-spectrum** communications to transmit a narrowband signal over a much larger bandwidth so that the spectral density of the signal in the channel looks like noise. The two different spread-spectrum techniques these tools employ are called *direct-sequence and frequency hopping*. The former hides information by phase-modulating the data signal (carrier) with a pseudorandom number sequence that both the sender and the receiver know. The latter divides the available bandwidth into multiple channels and hops between these channels (also triggered by a pseudorandom number sequence).

- **Echo hiding,** a form of data hiding, is a method for embedding information into an audio signal. It seeks to do so in a robust fashion, while not perceivably degrading the host signal (cover audio). Echo hiding introduces changes to the cover audio that are characteristic of environmental conditions rather than random noise, thus it is robust in light of many lossy data compression algorithms.

Like all good Steganographic methods, echo hiding seeks its data into data stream with minimal degradation of the original data stream. By minimal degradation, we mean that the change in the cover audio is either imperceivable or simply dismissed by the listener as a common non-objectionable environmental distortion.

The particular distortion we are introducing is similar to the resonances found in a room due to walls, furniture, etc. The difference between the stego audio and the cover audio is similar to the difference between listening to a compact disc on headphones and listening to it form speakers. With the headphones, we hear the sound as it was recorded. With the speakers, we hear the sound plus echoes caused by room acoustics. By correctly choosing the distortion we are introducing for echo hiding, we can make such distortions indistinguishable from those a room might introduce in the above speaker case.

### 3.7 Concealing Messages in Image and Audio Files Using S-Tools

S-Tools (Steganography Tools) brings you the capability of concealing files within various forms of data. Users of S-Tools can opt to encrypt their information using the strongest state-of-the-art encryption algorithms currently known within the academic world, so that even an enemy equipped with a copy of S-Tools cannot be completely sure data is hidden unless he has your secret passphrase.

You could use S-Tools to conceal private or confidential information that you don't want to fall into the wrong hands. You could use it to send information to another individual via a broadcast network such as Usenet. By agreeing on a passphrase you can keep the information out of unauthorised hands. Alternatively you could use S-Tools to verify your copyright over an image by storing an encrypted copyright statement in the graphic and extracting it in the event of a dispute.

- **How S-Tools hides your data**

S-Tools can hide multiple files in one object. If you have selected compression then the files are individually compressed and stored together with their names. If you are not using compression then just the raw file data is stored along with the names. Then S-Tools prepends some random garbage on to the front of the data in order to prevent two identical sets of files encrypting the same. The whole lot is then encrypted using the passphrase that you chose to generate the key (actually, MD5 is used to hash the passphrase down to 128 evenly distributed key bits). The encryption algorithms all operate in Cipher Feedback Mode (CFB).

It would be too easy to hide the data by just spreading it across the available bits in a linear fashion, so S-Tools seeds a cryptographically strong pseudo-random number generator from your passphrase and uses its output in order to choose the position of the next bit from the cover data to use.

For instance, if your sound file had 100 bits available for hiding, and you wanted to hide 10 bits in it, then S-Tools would not choose bits 0 through 9 as that would be trivially detectable by a potential enemy. Instead it might choose bits 63, 32, 89, 2, 53, 21, 35, 44, 99, 80. Or it might choose any ten others, it all depends on the passphrase that you enter. As you can see, the job of a potential enemy has just become very difficult indeed.

- **How data is hidden in sounds**

Sound samples are, by their very nature, inaccurate estimates of the correct value of the sound wave at a particular moment in time. The sound samples in Windows WAV files are stored as either 8 or 16 bit values that eventually get passed to the DA converter in your soundboard. For 8 bit samples this means that the values can range between 0 and 255. 16 bit samples range between 0 and 65535.

All S-Tools does is to distribute the bit-pattern that corresponds to the file that you want to hide across the least significant bits of the sound sample. For example, suppose that a sound sample had the following eight bytes of information in it somewhere:

132    134    137    141    121    101
               74    38

In binary, this is:

10000100 10000110 10000101 10001101
01111001 01100101 01001010 00100110
(LSB of each byte shown in italics)

Suppose that we want to hide the binary byte 11010101 (213) inside this sequence. We simply replace the LSB (Least Significant bit) of each sample byte with the corresponding bit from the byte we are trying to hide. So the above sequence will change to:

133    135    136    141    120    101    74
                       39

In binary, this is:

10000101 10000111 10001000 10001101
01111000 01100101 01001010 00100111

As you can clearly see, the values of the sound samples have changed by, at most, one value either way. This will be inaudible to the human ear, yet we have concealed 8 bits of information within the sample. This is the theory behind how S-Tools does its job.

- **How data is hidden in pictures**

All computer-based pictures are composed of an array of dots, called pixels, that make up a very fine grid. Each one of these pixels has its own colour, represented internally as separate quantities of red, green and blue. Within Windows, each of these colour levels may range between 0 (none of the colour) and 255 (a full amount of the colour). A pixel with an RGB value of 0 0 0 is black, and one with a value of 255 255 255 is white.

S-Tools works by 'spreading' the bit-pattern of the file that you want to hide across the least-significant bits (LSB's) of the colour levels in the image.

For a 24 bit image this is simple because 24 bit images are stored internally as RGB triples, and all we need to do is spread our bits and save out the new file. The drawback to this is that 24 bit images are uncommon at the moment, and would therefore attract the attention of those whose attention you are trying to avoid attracting! They are also very large as they contain 3 bytes for every pixel (for a 640x480 image this is 640x480x3=921600 bytes).

It is considerably more difficult to hide anything within a 256-colour image. This is because the image may already have over 200 colours which our meddling will carry to way over the absolute maximum of 256.

Looking at a little theory it is easy to see that an image with 32 or less colours will never exceed 256 colours, no matter how much we meddle with it. To see this, visualise the 3 LSB's of an RGB triple as a 3-bit number. As we pass through it in our hiding process we can change it to any one of 8 possible values, the binary digits from 000 to 111, one of which is the original pattern.

If one colour can 'expand' to up to 8 colours, how many distinct colours can we have before we are in danger of exceeding the limit of 256? Simple, 256/8=32 colours. There is no guarantee that 32 colours is our upper limit for every file that you want to hide though. If you're lucky the file will not change a colour to all of its 8 possible combinations and then we are able to keep one more of the original colours. In practice, however, you will often find pictures being reduced to the minimum of 32 colours. S-Tools tries to reduce the number of image colours in a manner that preserves as much of the image detail as possible.

I used a program called FileRay, to compare the binary values of two image files that were operated on using

S-Tools. cov.gif was used as the cover medium to hide stego.gif. The resultant file was named hidden.gif. Comparison of the original image to the stego data reveals changes in the LSB. Figure 11 shows the comparison between the two image files. The original image is shown in the bottom pane and the stego image is reflected in the top pane.

## 4. Steganalysis

As the techniques to hide information get more complicated and computationally involved, the detection of such cover medium has become considerably more challenging as well. However, given time, dedication and technology it is possible to detect the presence of hidden information in some stego mediums. A few tools have known signatures that may predict the presence of hidden information. Techniques like encryption and compression are used to make it difficult to decipher the hidden information. However, knowing the fact that there is hidden information present in the cover destroys the purpose of steganography.

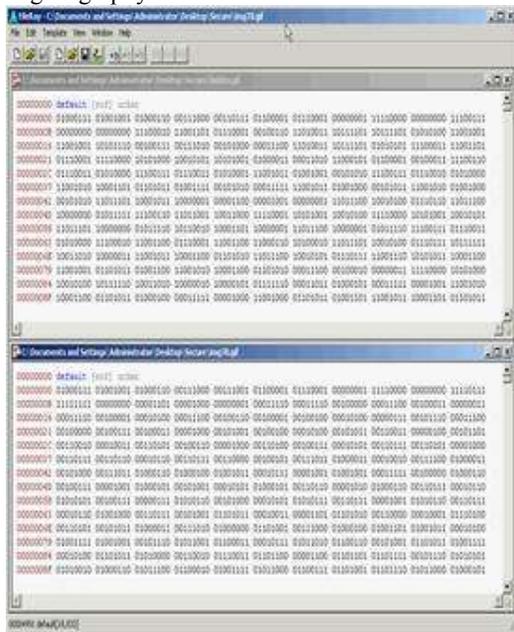

Figure 11: Comparison of cover image with stego in binary

Steganalysis is "the practice of attacking Steganographic methods by detection, destruction, extraction or modification of embedded data". This is the Steganographic analogue to cryptanalysis, which refers to attempts to break cryptographic protocols. With Cryptographic protocols, cryptanalysis is generally considered to be succeful if the adversary can retrieve the encrypted message. Steganography adds the additional requirement that the steganographically hidden message is not even detectable by the adversary; that is, not only should the attacker not be able to find the message, but he should not even know it exists. The definition of success in steganalysis depends upon your intent. For the security professional charged with protecting his employer's data, a successful result would be proving the existence of hidden data being sent, and not necessarily the ability to extract it. For the data thief, wishing to perhaps use a digital image that contains a protective watermark, success would be not only detecting the existence of the watermark, but would also require destroying it without damaging the integrity of the desired cover file.

Research shows that some well-known tools like S-Tools have known signatures and can be recognized if proper techniques are used. S-Tools works by reducing the number of colors of the cover image to 32, but expands them over several color palette entries, if the palette is then sorted by luminance, blocks of colors appear to be the same, but actually have a one-bit variance. This type of variance pattern is extremely rare in a natural image.

There are six formal categories of detection techniques available for steganalysis. The following table summarizes what the attacker has available to him in each case:

|  | Stego Object | Original Cover Object | Hidden Message | Stego Algorithm or Tool |
|---|---|---|---|---|
| **Stego only** | X |  |  |  |
| **Known cover** | X | X |  |  |
| **Known message** | X |  | X |  |
| **Chosen stego** | X |  |  | X |
| **Chosen message** | X |  |  |  |
| **Known stego** | X | X |  | X |

A *stego only attack*, while considered the most difficult attack in that one has the least information to go on, is far from impossible, especially if ones goal is to merely detect that there is a hidden message and not necessarily have the need to extract it. For text files using empty space methods, like the one used with the Snow tool, merely opening the document with an editor that shows formatting codes would indicate that there were oddities in the formatting. The message could be easily destroyed through simply removing the extra spaces and tabs. Similarly, for other text methods such as line and word shift methods, visual inspection of the text itself could indicate anomalies. For image and audio files with messages embedded with LSB methodology, detection with only the stego file available is a little more difficult. Detection in this case would rely on the appearance of visual or audible distortions or patterns.

With the "Known-cover" method, one has both the original innocent cover file as well as the resulting stego file. Anomalous patterns and excess noise in the stego file are much more easily detectable when comparing it to the original, particularly if the file format makes use of compression (as in JPEG files) that would show up as excess noise even in as innocent file. Destruction or distortion of LSB encoded messages in images in image files can be simply a matter of zeroing out the LSB fields of the file in question, image conversion, cropping or the application of other image formatting changes. For audio files, methods to damage the hidden message include then introduction of a random relative amplitude signal and reconstructing the file by ignoring bad signals.

A "known-message" attack gives the attacker the knowledge of the secret message that is hidden in the file.

A "chosen-stego" attack provides the attacker with the extraction tool to reach the data, and the "chosen message" technique assumes the attacker has the stego tool itself, and can embed and detect message at will.

There are other ways to break up attack-types, and these are also useful in describing the vulnerabilities of various methods. Wayner divides common attack methods by functional properties rather than adversarial assumptions; attacks are divided into visual or aural attacks, structural attacks, and statistical attacks. Visual and aural attacks describe the human factor in attacks; humans can often perceive the modifications in the cover object because it doesn't look or sound right. In text steganography this can be extended to format-based, lexical, grammatical, semantic, and rhetorical attacks. Among others. Structural attacks refer to detecting the patterns in modifications made in the data format (for example, using extra space in files or encoding schemes to store information is often detectable through structural attacks). Statistical attacks detect anomalies in the statistical profile of the stego-object (for example, images whose color palette has been changed to hide information often contain non-standard colors or ranges of colors which would not normally be generated by image software).

## 5. Digital Watermarking -- Steering the Future of Security

Watermarks were first used in Europe to identify the guild that manufactured paper. They were like trademarks or signatures. Varying the paper's density creates watermarks in paper. Normally invisible, a watermark image becomes visible as darker and lighter areas when the paper is held up to the light. Wire or relief sculptures are placed in the paper mold and when the paper slurry is drained of its water and dried the thinner areas created by the wire or sculpture show clearly when held up to the light. Watermarks are still used in quality stationary and have even been added to currencies of various countries.

A watermark is a form, image or text that is impressed onto paper, which provides evidence of its authenticity. Digital watermarking is an extension of this concept in the digital world. In recent years the phenomenal growth of the Internet has highlighted the need for mechanisms to protect ownership of digital media. Exactly identical copies of digital information, be it images, text or audio, can be produced and distributed easily. In such a scenario, who is the artist and who the plagiarist? It's impossible to tell--or was, until now. Digital watermarking is a technique that provides a solution to the longstanding problems faced with copyrighting digital data.

Digital watermarks are pieces of information added to digital data (audio, video, or still images) that can be detected or extracted later to make an assertion about the data. This information can be textual data about the author, its copyright, etc; or it can be an image itself. The digital watermarks remain intact under transmission / transformation, allowing us to protect our ownership rights in digital form.

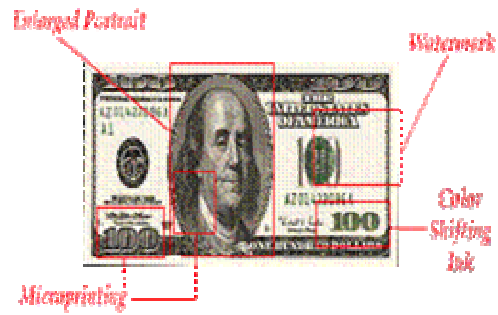

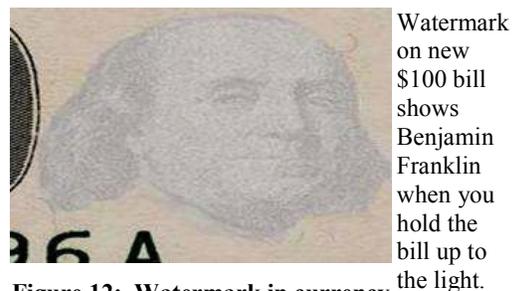

Watermark on new $100 bill shows Benjamin Franklin when you hold the bill up to the light.

**Figure 12: Watermark in currency**

A given watermark may be unique to each copy (e.g. to identify the intended recipient), or be common to multiple copies (e.g. to identify the document source).

In either case, the watermarking of the document involves the transformation of the original into another form. This distinguishes digital watermarking from digital fingerprinting, where the original file remains intact and a new created file 'describes' the original file's content.

### 5.1 General Framework for Watermarking

A digital watermark is, in essence, a hidden message, within a digitized image, video or audio recording. The watermark is integrated into the content itself. So it requires no additional storage space.

In general, any watermarking scheme (algorithm) consists of three parts.

- The watermark
- The encoder (insertion algorithm)
- The decoder and comparator (verification or extraction or detection algorithm)

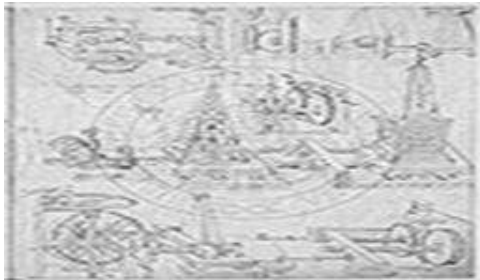

**Figure 13: Digital Copy of fifteenth century drawing with digital watermark superimposed.**

Each owner has a unique watermark or an owner can also put different watermarks in different objects. The marking algorithm incorporates the watermark into the object. The verification algorithm authenticates the object determining both the owner and the integrity of the object.

#### 5.1.1 Encoding Process
Let us denote an image by I, a signature by S=*s1,s2,…* and the watermarked image by I. E is an encoder function, it takes an image I and a signature S, and it generates a new image which is called watermarked image I, mathematically,

$$E(I,S) = I' \quad \text{............ (1)}$$

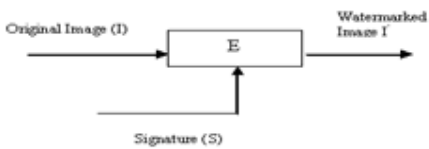

**Figure 14: Encoder**

#### 5.1.2 Decoding Process
A decoder function D takes an image J (J can be a watermarked or un-watermarked image, and possibly corrupted) whose ownership is to be determined and recovers a signature S' from the image. In this process an additional image I can also be included which is often the original and un-watermarked version of J. This is due to the fact that some encoding schemes may make use of the original images in the watermarking process to provide extra robustness against intentional and unintentional corruption of pixels. Mathematically,

$$D(J,I) = S' \quad \text{............ (2)}$$

The extracted signature S' will then be compared with the owner signature sequence by a comparator function $C_\delta$ and a binary output decision generated. It is 1 if there is match and 0 otherwise, which can be represented as follows.

$$C_\delta(S', S) = \begin{cases} 1, & c <= \delta \\ 0, & \text{Otherwise} \end{cases}$$

Where C is the correlator, x= $C_\delta$ (S', S). c is the correlation of two signatures and $\delta$ is certain threshold. Without loss of generality, watermarking scheme can be treated as a three-tupple (E, D, $C_\delta$). Following figures demonstrate the decoder and the comparator.

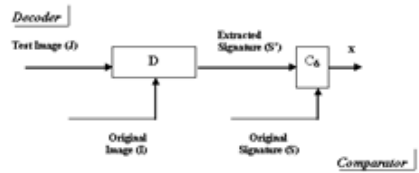

**Figure 15: Decoder**

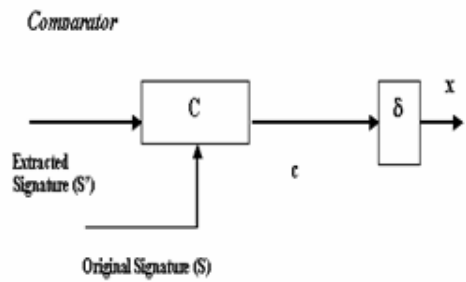

**Figure 16: Comparator**

A watermark must be detectable or extractable to be useful. Depending on the way the watermark is inserted and depending on the nature of the watermarking algorithm, the method can involve very distinct approaches. In some watermarking schemes, a watermark can be extracted in its exact form, a procedure we call watermark extraction. In other cases, we can detect only whether a specific given watermarking signal is present in an image, a procedure we call watermark detection. It should be noted that watermark extraction can prove ownership whereas watermark detection can only verify ownership.

## 5.2 Watermarking Algorithms

Watermarks and watermarking techniques can be divided into various categories in various ways. The watermarks can be applied in spatial domain. An alternative to spatial domain watermarking is frequency domain watermarking. It has been pointed out that the frequency domain methods are more robust than the spatial domain techniques.

- **A simple Spatial watermarking algorithm -- The LSB technique**

The LSB technique is the simplest technique of watermark insertion. If we specifically consider still images, each pixel of the color image has three components -- red, green and blue. Let us assume we allocate 3 bytes for each pixel. Then, each colour has 1 byte, or 8 bits, in which the intensity of that colour can be specified on a scale of 0 to 255.

So a pixel that is bright purple in colour would have full intensities of red and blue, but no green. Thus that pixel can be shown as
$$X0 = \{R=255, G=0, B=255\}$$
Now let's have a look at another pixel:
$$X1 = \{R=255, G=0, B=254\}$$

We've changed all the value of B here. But how much of a difference does it make to the human eye? For the eye, detecting a difference of 1 on a color scale of 256 is almost impossible. Now since each color is stored in a separate byte, the last bit in each byte stores this difference of one. That is, the difference between values 255 and 254, or 127 and 126 is stored in the last bit, called the Least Significant Bit (LSB).

Since this difference does not matter much, when we replace the color intensity information in the LSB with watermarking information, the image will still look the same to the naked eye. Thus, for every pixel of 3 bytes (24 bits), we can hide 3 bits of watermarking information, in the LSBs.

Thus a simple algorithm for this technique would be:

```
Let W be watermarking information
For every pixel in the image, Xi
Do Loop:
    Store the next bit from W in the
    LSB position of Xi [red] byte
    Store the next bit from W in the
    LSB position of Xi [green] byte
    Store the next bit from W in the
    LSB position of Xi [blue] byte
End Loop
```

To extract watermark information, we would simply need to take all the data in the LSBs of the color bytes and combine them. Image manipulations, such as resampling, rotation, format conversions and cropping, will in most cases result in the watermark information being lost.

- **Frequency based Watermarking**

Watermarking in the frequency domain involves selecting the pixels to be modified based on the frequency of occurrence of that particular pixel. This is to overcome the greatest disadvantage of techniques operating in the spatial domain i.e. susceptibility to cropping. The mosaic attack (In a mosaic attack, the attacker breaks up the entire watermarked image into many small parts. For example, a watermarked image on a web page can be cut up and reassembled as a whole using tables in HTML. The only defence against this attack is to tile a very small watermark all over the image, and allow retrieval of the watermark from any of the small subsections of the fragmented image. However, the attacker can always create smaller blocks, and the watermarked image also has to be large enough to be distinguishable), defeats most implementations of digital watermarking operating in the spatial domain but the frequency domain watermarking is less susceptible.

The LSB technique can also be applied in the frequency domain selecting the pixels according to frequency, though not robust. Common transforms, such as Fast Fourier Transforms, alter the value of pixels within the original image based on their frequencies. The watermark is more commonly applied to the lower frequencies within an image as higher frequencies are usually lost when an image is compressed or to frequencies considered to contain perceptually significant information. Frequency based techniques result in a watermark that is dispersed throughout the image, therefore, less susceptible to attack by cropping. However these techniques are susceptible to standard frequency filters and lossy compression algorithms, which tend to filter out less significant frequencies.

## 5.3 Types of Digital Watermarks

Watermarks and watermarking techniques can be divided into various categories in various ways. The watermarks can be applied in spatial domain. An alternative to spatial domain watermarking is frequency domain watermarking.

Visible watermark is a secondary translucent overlaid into the primary image. The watermark appears visible to a casual viewer on a careful inspection.

A fragile watermark is a mark, which is sensitive to a modification of the stego-medium. A fragile watermarking scheme should be able to detect any change in the signal and identify where it has taken place and possibly what the signal was before modification. It serves at proving the authenticity of a document.

On the opposite, a robust watermark should be stuck to the document it has been embedded in, in such a way that any signal transform of reasonable strength cannot remove the watermark. Hence a pirate willing to remove the watermark will not succeed unless they debase the document too much to be of commercial interest.

Dual watermark is a combination of a visible and an invisible watermark. In this type of watermark an invisible watermark is used as a back up for the visible watermark as clear from the following diagram.

Private watermarking and non-blind-watermarking mean the same: the original cover signal is required during the detection process. By asymmetric watermarking or public-key watermarking, people refer to watermarking schemes with properties reminding asymmetric cryptosystem (or public key cryptosystem). No such system really exists yet although some possible suggestions have been made. In this case, the detection process (and in particular the detection key) is fully known to anyone as opposed to blind watermarking where a secret key is required. So here, only a 'public key' is needed for verification and a 'private key' (secret) is used for the embedding though. Knowledge of the public key does not help to compute the private key, it does not either allow removal of the mark nor it allows an attacker to forge a mark.

Source-based watermark are desirable for ownership identification or authentication where a unique watermark identifying the owner is introduced to all copies of a particular image being distributed. A source-based watermark could be used for authentication and to determine whether a received image or other electronic data has been tempered with. The watermark could also be estimation-based where each distributed copy gets a unique watermark identifying the particular buyer. The destination-based watermark could be used to trace the buyer in the case of illegal reselling.

## 5.4 Applications of Digital Watermarks

**Visible Watermark**
Visible watermarks can be used in the following cases:

- Visible watermarking for enhanced copyright protection. In such situations, where images are made available through Internet the content owner is concerned that the images will be used commercially (e.g. Imprinting coffee mugs) without payment of royalties. Here the content owner desires an ownership mark, that is visually apparent, but which does not prevent image being used for other purposes (e.g. scholarly research).
- Visible watermarking used to indicate ownership origins. In this case images are made available through the Internet and the content owner desires to indicate the ownership of the underlying materials (library manuscript), so an observer might be encouraged to patronize the institutions that own the material.

**Invisible Robust Watermark:**
Invisible robust watermarks find application in following cases.

- Invisible watermarking to detect misappropriated images. In this scenario, the seller of digital images is concerned, that his, fee generating images may be purchased by an individual who will make them available for free, this would deprive the owner of licensing revenue.
- Invisible watermarking as evidence of ownership. In this scenario, the seller that of the digital images suspects one of his images has been edited and published without payment of royalties Here, the detection of the seller's watermark in the image is intended to serve as evidence that the published image is property of seller.

**Invisible Fragile Watermarks**
Following are the applications of invisible fragile watermarks.

- Invisible watermarking for a trustworthy camera. In this scenario, images are captured with a digital camera for later inclusion in news articles. Here, it is the desire of a news agency to verify that an image is true to the original capture and has not been edited to falsify a scene. In this case, an invisible watermark is embedded at capture time; its presence at the time of publication is intended to indicate that the image has not been attended since it was captured.
- Invisible watermarking to detect alternation of images stored in a digital library. In this case, images (e.g. human fingerprints) have been scanned and stored in a digital library; the content owner desires the ability to detect any alternation of the images, without the need to compare the images to the scanned materials.

# 6. Audio Watermarking

Digital audio watermarking involves the concealment of data within a discrete audio file. Applications for this technology are numerous. Intellectual property protection is currently the main driving force behind research in this area. To combat online music piracy, a digital watermark could be added to all recording prior to release, signifying not only the author of the work, but the user who has purchased a legitimate copy. Newer operating systems equipped with digital rights management software (DRM) will extract the watermark from audio files prior to playing them on the system. The DRM software will ensure that the user has paid for the song by comparing the watermark to the existing purchased licenses on the system.

- **DC Watermarking Scheme**

This section details the implementation of a digital audio watermarking scheme, which can be used to hide auxiliary information within a sound file.

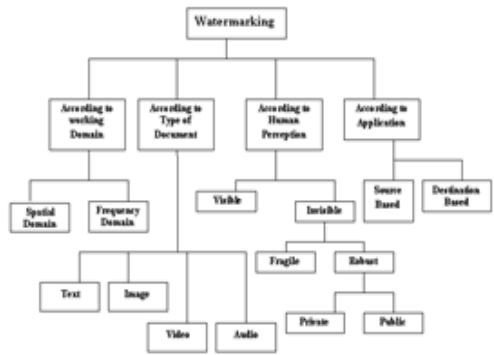

Figure 17: Types of Watermarking Techniques

The DC watermarking scheme hides watermark data in lower frequency components of the audio signal, which are below the perceptual threshold of the human auditory system.

❖ **Watermark Insertion**

The process of inserting a digital watermark into an audio file can be divided into four main processes (see Figure 8). A original audio file in wave format is fed into the system, where it is subsequently framed, analyzed, and processed, to attach the inaudible watermark to the output signal.

**Framing**
The audio file is portioned into frames which are 90 milliseconds in duration. With a 90 ms frame size, our bit rate for watermarked data is equal to 1 / 0.09 = 11.1 bits per second.

**Spectral Analysis**
Next, spectral analysis is performed on the signal, consisting of a fast Fourier transform (FFT), which allows us to calculate the low frequency components of each frame, as well as the overall frame power. From the FFT, we are now able to determine the low frequency (DC) component of the frame as well as the frame spectral power.

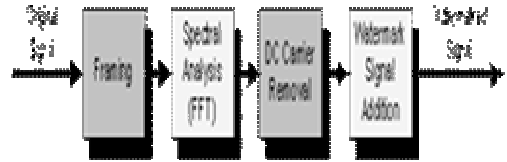

**Figure 18. Watermark Insertion Process.**

**DC Removal**
From the above spectral analysis of each frame, we have calculated the low frequency (DC) component F(1), which can now be removed by subtraction from each frame.

**Watermark Signal Addition**
From the spectral analysis completed previously, we calculated the spectral power for each frame, which is now utilised for embedding the watermark signal data. The power in each frame determines the amplitude of the watermark which can be added to the low frequency spectrum.

❖ **Watermark Extraction**
The process of extracting the digital watermark from the audio file is similar to the technique for inserting the watermark. The computer processing requirements for extraction are slightly lower. A marked audio file in wave format is fed into the system, where it is subsequently framed, analysed, and processed, to remove the embedded data which exists as a digital watermark.

**Framing**
As with the insertion process, the audio file is partitioned into frames which are 90 milliseconds in duration.

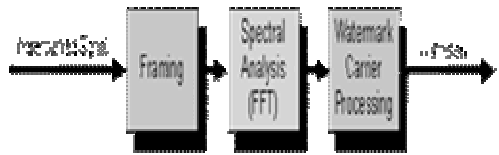

**Figure 19: Watermark Extraction Process**

**Spectral Analysis**
Subsequent to the framing of the watermarked audio signal, we perform spectral analysis on the signal, consisting of a fast Fourier transform (FFT), which again allows us to calculate the low frequency components of each frame, as well as the overall frame power.

**Watermark Signal Extraction**

From the spectral analysis completed previously, we calculated the spectral power for each frame, which allows us to examine the low frequency power in each frame and subsequently extract the watermark.

In order to attain higher hidden data density in the watermarked signal, more advanced techniques must be used such as spread spectrum, phase encoding, or echo hiding.

# 7. Conclusions

In this tutorial, we take an introductory look at information hiding techniques. Historical detail is discussed. Several methods for hiding data in text, image, and audio are described, with appropriate introductions to the environment of each medium, as well as the strengths and weaknesses of each method. Most data hiding systems take advantage of human perceptual weaknesses, but have weaknesses of their own. In areas where cryptography and strong encryption are being outlawed, citizens are looking at steganography to circumvent such policies and pass messages covertly. Commercial applications of steganography in the form of digital watermarks are currently being used to track the copyright and ownership of electronic media. We conclude that for now, it seems that no system of data hiding is totally immune attack. However, steganography has its place in security. It in no way can replace cryptography, but is intended to supplement it. Its application in watermarking for use in detection of unauthorised, illegally copied material is continually being realised and developed.

*Courtesy:* **WWW**